# PHYSICS IN DISCRETE SPACES (D): THE STANDARD MODEL OF PARTICLES AND BEYOND

Pierre Peretto
Laboratoire de Physique et Modélisation des Milieux Condensés
CNRS LPMMC Grenoble (France)
peretto.pierre@neuf.fr

Abstract:
We show that the model of discrete spaces that we have proposed in previous contributions gives a comprehensive and detailed interpretation of the properties of the standard model of particles. Moreover the model also suggests the possible existence of a new family of particles.

## 1. INTRODUCTION

According to the model of discrete space-time that we have put forward in preceding contributions [1], [2], [3] the universe as a whole could be considered as a sort of spin glass that is a set of randomly interacting Ising spins called *cosmic bits*. This model accounts for several fundamental properties of the physical world. To be convinced however that the model also provides a correct description of natural phenomena in general something essential is still missing: It must explain the properties of matter itself that is it must provide an interpretation of the standard model of particles. The model of discrete space-time offers this possibility because in a discrete space a physical point, called a *world point*, has a finite physical size and is not identified with a mathematical point. In the usual interpretation of natural phenomena a physical point, that is a mathematical point, is a system devoid of any structure. A discrete space-time model allows, on the contrary, an internal organization to develop inside world points, providing a way for the building of more or less complex structures. This is the object of the present, and final, contribution.

## 2. THE STRUCTURE OF ELEMENTARY PARTICLES IN DISCRETE SPACES

The hypothesis that the universe is discrete suggests that the elementary particles of the standard model are structures made of world points with a given symmetry. Let us recall that the cosmic bits of a world point are all connected to each other through ferromagnetic interactions [1]. Then the Minkowski metrics disappears inside a world point which therefore seems punctual from a physical point of view. According to [2] the size $l^*$ of a world point would be of the order of $l^* \cong 0.5\times10^{-21}$ cm.

Our approach suggests that a particle could be structured along a three levels organization

i)- A particle is built around a special world point (whose size is $l*$), called its *seed*. The symmetry properties of the seed determine the symmetry of the whole particle.
ii)- The seed modifies the polarization states and metric matrices of the neighbouring world points over a range $\rho l*$. This region of influence is called the *core* of the particle and the system made of a seed and its core constitutes a bare particle
iii)- Finally the bare particle is surrounded by a cloud of gauge particles whose nature and properties are determined by the symmetry of the seed according to the gauge symmetry mechanisms that have been discussed in [2]. The system made of a bare particle and its cloud of virtual particles constitutes a dressed particle.

## 3. THE MASSES OF BARE PARTICLES

In our model of discrete spaces the Lagrangian $\Lambda(\psi)$ of a physical system comprised of *N* world points is given by

$$\Lambda(\psi) = \psi^T (\Delta \otimes G)\psi \qquad (1)$$

where

-$\psi$ is a state of the system, $\psi = \{\phi_i\}$ and $\phi_i$ is the 4-dimensional polarization state of world point *i* $(i = 1,...,N)$.

-$\Delta$ is a $(N \times N)$, symmetric, random matrix that can be factorized along $\Delta = D^T D$ . *D* is a *N*-dimensional triangular random matrix ($D_{i,j<i} = 0$), and $D^T$ is its transpose. The elements $\Delta_{ij}$ of $\Delta$ describe the interactions between world points *i* and *j*, $(i, j = 1,..., N)$

-*G* is a set $\{G_i\}$ of *N* $4\times 4$ metric matrices $G_i$ whose elements describe the interactions between the components $\varphi_{i\mu}$ and $\varphi_{i\nu}$ of the polarization state $\phi_i = \{\varphi_{i\mu}\}$ $(\mu = 1,...,4)$ of world point *i*.

-the suffix "*T*" is for hermitian conjugation.

In vacuum, that is in a system devoid of particles, the $G_i$ matrices are all equal to each other: $G_i = G_0$ where $G_0$ is the (Lorentzian) metric matrix of vacuum [1].

$$G_0 = \begin{pmatrix} -\frac{1}{c^2} & & & \\ & 1 & & \\ & & 1 & \\ & & & 1 \end{pmatrix}$$

c is a dimensionless constant given by the ratio of length to time standards [1]. A system contains a particle *P* if one of the metric matrices, that of world point *i*, is such that $G_i = G_P$ (and $G_j = G_0$ for $j \neq i$) where $G_P$ transforms according to the symmetry that characterizes *P*. A physically possible state of such a quantum system is obtained by minimizing $\Lambda(\psi)$ under the constraint $\psi^T\psi = 1$. This gives the following eigenvalue equation

$$(\Delta \otimes G)\psi_P = \kappa_P \psi_P\,. \qquad (2)$$

The eigenvalue $\kappa_P$ is a Lagrange multiplier. We have seen in [1] and [3] that equation (2) is in fact a Klein-Gordon equation

$$\left(-\frac{1}{c^2}\frac{\partial^2}{\partial t^2}+\Delta\right)\psi_P=\left(\frac{cm_P}{\hbar}\right)^2\psi_P \qquad (3)$$

Therefore the mass of the particle $P$ is given by

$$m_P=\frac{\hbar}{c}(\kappa_P)^{1/2} \qquad (4)$$

which translates into the mass $m_P$ being proportional to the square root of the eigenvalue $\kappa_P$. An eigenstate $\phi_s$ of an isolated seed is determined by the following eigenstate equation:

$$G_P\phi_s=\kappa_s\phi_s$$

with "*s*" standing for seed. The mass of the seed can be defined from the eigenvalue $\kappa_s$ by

$$m_s=\frac{\hbar}{c}(\kappa_s)^{1/2} \qquad (5)$$

We show in Appendix A that the symmetry of the core, the bare particle, and the dressed particle as a whole is that of the seed itself. Moreover the eigenvalue associated with the core relative to vacuum is proportional to $\kappa_s$ which means that the eigenvalue associated with the bare particle is also proportional to $\kappa_s$ :

$$\kappa_P^b=K_S\kappa_S$$

where "*b*" is for bare and $K_S$ is defined in Appendix A. The mass of a bare particle finally writes

$$m_P^b=\left(\frac{\hbar}{c}\right)(K_s\kappa_s)^{1/2} \qquad (6)$$

The parameter $K_s$ only depends on the symmetry of the seed: $K_s=K_Q$ for quarks (symmetry SU(3)) and $K_s=K_L$ for leptons (symmetry U(1)). To obtain the mass of dressed particles one must add the effects of clouds of virtual bosons which, in fact, constitute the main part of experimental masses.

This way of introducing particle masses seems to go against the usual conclusion that all particles must have zero masses due to the dichotomy of the universe between a right and a left universe. The usual argument that leads to that conclusion is reminded in Appendix B. The universe dichotomy is introduced to account for the experimental observation that the neutrinos are left handed. We show in section 8 that this property of neutrinos is, in reality, a consequence of the model and no Higgs mechanism is needed to give a mass to particles. It must be stressed, however, that a Higgs scalar field, identified with the world point polarization state $\phi_i$, is still necessary to give them an existence.

## 4. SYMMETRY AS AN ORGANIZING PRINCIPLE OF PARTICLES

A finite group, the permutation group $S_4$ of four objects, plays a prominent role in our approach as already stressed in [1]. Physics, indeed, must be invariant under any permutation of the four axes used to describe the internal states $\phi_P$ of world point *i* since no metrics can be defined inside the world point. The permutation group $S_4$ of four objects has 24 elements distributed into 5 classes. The table of characters of $S_4$, already given in [1], is repeated in Table I

| classes | $(1):1$ | $(ab):6$ | $(ab)(cd):3$ | $(abc):8$ | $(abcd):6$ |
|---|---|---|---|---|---|
| $\Gamma_1$ | 1 | 1 | 1 | 1 | 1 |
| $\Gamma_1^*$ | 1 | $-1$ | 1 | 1 | $-1$ |
| $\Gamma_2$ | 2 | 0 | 2 | $-1$ | 0 |
| $\Gamma_3$ | 3 | 1 | $-1$ | 0 | $-1$ |
| $\Gamma_3^*$ | 3 | $-1$ | $-1$ | 0 | 1 |

Table-I: Table of characters of $S_4$

The matrix $G_P$ must commute with any four dimensional representation of group $S_4$ and therefore it must transform according to direct sums of irreducible representations of $S_4$. There are three types of irreducible representations. On one hand $\Gamma_1$ , and $\Gamma_3$ are insensitive to mirror operations that is to an odd number of permutations. On the other $\Gamma_1^*$ and $\Gamma_3*$ are sensitive to mirror operations. Finally $\Gamma_2$ does not take the mirror operations into account.

Let us try to build the matrix $G_P$ by using $\Gamma_1$, $\Gamma_2$ and $\Gamma_3$.

The problem of finding four-dimensional matrices $G_P$ as direct sums of $\Gamma_1$, $\Gamma_2$, and $\Gamma_3$ representations has the following solutions.

i)- 4=1+3
ii)- 4=2+2
iii)- 4=1+1+2
iv)-4=1+1+1+1

i- The first possibility is to build $G_P$ as a direct sum of $\Gamma_1$ and $\Gamma_3$

$$G_B \approx \Gamma_1 \oplus \Gamma_3$$

This corresponds to solution (i) (4=1+3).

We say that world points that are polarized along this symmetry display a *bosonic polarization*. In its diagonal form $G_B$ writes accordingly

$$G_B = \begin{pmatrix} \kappa_{B1} & & & \\ & \kappa_{B2} & & \\ & & \kappa_{B2} & \\ & & & \kappa_{B2} \end{pmatrix}$$

The analytical forms of matrices $G_P$ may be found from the dynamics of their corresponding particles. The dynamics of boson fields follows a Klein-Gordon equation (3) which, as we have seen, is the translation of the discrete eigenvalue equation

$$(\Delta \otimes G_B)\psi = (D^T D \otimes G_B)\psi = \kappa\psi$$

in continuum space.

The matrix $G_B$ is obtained from the lagrangian $\Lambda_B = \Delta \otimes G_B$ by isolating a world point *i* from its neighbourhood which is achieved by letting $\Delta_{ij} = \delta_{ij}$ that is by replacing the differential operators $\partial_\mu^2$ by Dirac functions $\partial_\mu^2 \to \delta(x_\mu)$ with $\delta(x_\mu)=1$ if $x_\mu = 0$ and $\delta(x_\mu)=0$ otherwise. $G_B$ is then given by

$$G_B = \begin{pmatrix} -1/c^2 & & & \\ & 1 & & \\ & & 1 & \\ & & & 1 \end{pmatrix}$$

$G_B$, therefore, is proportional to the metric matrix of vacuum $G_0$. The eigenvalues of $G_B$ are $\kappa_{B1} = -1/c^2$, a non degenerate time-like eigenvalue and $\kappa_{B2} = 1$, a three fold degenerate space-like eigenvalue. The ground state of $G_B$ (the vacuum state $\phi_0$ of the physical system) is the eigenstate associated with the eigenvalue $\kappa_{B1}$:

$$\phi_0 = \begin{pmatrix} 1 \\ 0 \\ 0 \\ 0 \end{pmatrix}$$

The vacuum state has one time and zero space components.
ii- The second possibility is to build $G_P$ on the representation $\Gamma_2$.

$$G_F \approx \Gamma_2 \oplus \Gamma_2 = 1^{(2)} \otimes \Gamma_2$$

This corresponds to solution (ii) (4=2+2)
We say that world points that are polarized along this symmetry display a *fermionic polarization*. The representation $\Gamma_2$ determines the fermionic, or spinor, character of the polarization. Since a fermionic polarization implies the representation $\Gamma_2$ twice, it is called a bispinor. In its diagonal form $G_F$ writes

$$G_F = \begin{pmatrix} \kappa_{F1} & & & \\ & \kappa_{F1} & & \\ & & \kappa_{F2} & \\ & & & \kappa_{F2} \end{pmatrix}$$

As we know, the dynamics of fermion fields, that is Dirac equation, is obtained by factorizing the Klein-Gordon equation. The Dirac equation writes

$$\left( \sum_\mu \mathrm{i}\hbar\gamma^\mu \partial_\mu - mc \right)\psi = 0$$

This equation may be obtained from the discrete eigenvalue equation

$$(D \otimes G_F)\psi = \kappa\psi$$

and the matrix $G_F$ is found by isolating a world point from its neighbourhood that is by letting $\partial_\mu \to \delta(x_\mu)$ in the lagrangian $\Lambda_F = D \otimes G_F$.
Therefore

$$G_F = \left[ \frac{\gamma^1}{c} + \gamma^2 + \gamma^3 + \gamma^4 \right],$$

where the Dirac matrices $\gamma^\mu$ are expressed in a basis (the Dirac representation) where the indices $\mu$ of matrices $\gamma^\mu$ are identified with the Lorentz indices $\mu$ of space-time itself. The matrix $\gamma^1$

$$\gamma^1 = \begin{pmatrix} 1 & & & \\ & 1 & & \\ & & -1 & \\ & & & -1 \end{pmatrix}$$

is introduced so as to make hermitian the operator $\Lambda_F$ .Then

$$G_F = \gamma^1\left(\frac{1}{c}\gamma^1 + \gamma^2 + \gamma^3 + \gamma^4\right) \quad (7)$$

and the matrix $G_F$ finally writes

$$G_F = \begin{pmatrix} 1/c & 0 & 1 & 1+i \\ 0 & 1/c & 1\text{-}i & -1 \\ 1 & 1+i & 1/c & 0 \\ 1\text{-}i & -1 & 0 & 1/c \end{pmatrix}$$

As expected $G_F$ has two real two-fold degenerate eigenvalues $\kappa_{F1} = \left(1/c - \sqrt{3}\right)$ and $\kappa_{F2} = \left(1/c + \sqrt{3}\right)$.

iii- The third possibility that apparently does not play any role in the standard model, is

$$G_{ns} \approx \Gamma_1 \oplus \Gamma_1 \oplus \Gamma_2$$

This corresponds to the solution (iii) (4=1+1+2). "*ns*" is for non standard. This solution will be further discussed in section V.

Let us see now how $G_B$ and $G_F$ may generate the fundamental particles of the standard model of particles.

Super-symmetry theory (Susy) puts forward that fermions and bosons may be considered as two aspects of the very same objects. In a spirit close to that of Susy it is suggested here that the fundamental particles are objects that associate world points with different symmetries. A bare particle would, accordingly, be made of a pair of coupled world points, one undergoing a fermionic polarization and the other a bosonic polarization [1]. The idea that an elementary particle could be formed of a couple of bosonic and fermionic sub-particles has also been suggested by Koide [4].

The state of a particle is then represented in a 16-dimensional vector space that is obtained by the direct product of the two 4-dimensional spaces associated with the members of the pair that form the particle. The states must therefore transform as

$$(\Gamma_1 \oplus \Gamma_3) \otimes (\Gamma_2 \oplus \Gamma_2)$$

that may be expanded along

$$(\Gamma_1) \otimes (\Gamma_2 \oplus \Gamma_2) \oplus (\Gamma_3) \otimes (\Gamma_2 \oplus \Gamma_2)$$

There are therefore two types of particles. On one hand the two particles associated with $(\Gamma_1) \otimes (\Gamma_2 \oplus \Gamma_2) = (\Gamma_1 \otimes \Gamma_2) \oplus (\Gamma_1 \otimes \Gamma_2)$ are called leptons. On the other hand the two particles associated with $(\Gamma_3) \otimes (\Gamma_2 \oplus \Gamma_2) = (\Gamma_3 \otimes \Gamma_2) \oplus (\Gamma_3 \otimes \Gamma_2)$. are called quarks. Since all these particles transform along $\Gamma_2$ they are all fermions. The 4-dimensional matrix that represents $(\Gamma_2 \oplus \Gamma_2)$ has two eigenvalues $\kappa_{F1}$ and $\kappa_{F2}$ and, therefore there are two sorts of leptons and two sorts of quarks. The lepton associated with the eigenvalue $\kappa_{F2} = (1/\mathrm{c} + \sqrt{3})$ is called the neutrino, that associated with $\kappa_{F1} = (1/\mathrm{c} - \sqrt{3})$ is called the electron. The quark associated with eigenvalue $\kappa_{F2} = (1/\mathrm{c} + \sqrt{3})$ is called the down quark *d* and, finally, the quark associated with eigenvalue $\kappa_{F1} = (1/\mathrm{c} - \sqrt{3})$ is called the up quark *u*. The representation $\Gamma_3$ then introduces a three-fold degeneracy, associated with colour quantum number, for the states of quarks.

Let us now consider how antiparticles may be derived from the same formalism. The different types of particles have been determined by using the mirror insensitive representations $\Gamma_1$ $\Gamma_2$ and $\Gamma_3$. The types of antiparticles are determined by using the mirror sensitive representations $\Gamma_1{}^*$, $\Gamma_3{}^* = \Gamma_1^* \otimes \Gamma_3$ and $\Gamma_2^* = \Gamma_1^* \otimes \Gamma_2 = \Gamma_2$. Since $\Gamma_1{}^*$ and $\Gamma_3{}^*$ are sensitive to mirror operations the state $\bar{\phi}$ associated with an anti-particle is obtained from the state $\phi$ associated with a particle through a mirror transformation that is a sign change of one of its components. Nothing, however, determines which component has to be modified and all signs have to be changed at once that is

$$\bar{\phi} = \begin{pmatrix} -\varphi_1 \\ -\varphi_2 \\ -\varphi_3 \\ -\varphi_4 \end{pmatrix} = \begin{pmatrix} -1 & & & \\ & -1 & & \\ & & -1 & \\ & & & -1 \end{pmatrix} \begin{pmatrix} \varphi_1 \\ \varphi_2 \\ \varphi_3 \\ \varphi_4 \end{pmatrix} = \mathrm{C}\phi$$

C is the charge conjugation operator.

The states of anti-particles then transform as

$$(\Gamma_1^* \oplus \Gamma_3^*) \otimes (\Gamma_2 \oplus \Gamma_2)$$

that may be expanded along

$$(\Gamma_1^*) \otimes (\Gamma_2 \oplus \Gamma_2) \oplus (\Gamma_3^*) \otimes (\Gamma_2 \oplus \Gamma_2)$$

Antiparticles are therefore organized exactly as particles. They have exactly the same properties as the corresponding particles, except for the sign of electric charges which is sensitive to charge conjugation. We observe that this interpretation of antiparticles avoids the introduction of a Dirac sea.

We have so far obtained a description of the organization of one family of elementary particles that exactly fits the standard model of particles. However there are three families of particles with identical properties except for their masses. The symmetry properties are the same for the three families and one must admit accordingly that the matrices $G_P$ are the same for the three families but while the particles of the first family are stable the particles belonging to the other families are unstable. Therefore, the existence of these families cannot be looked for in some new type of symmetry similar to SU(3) for example. An alternative interpretation is proposed below.

## 5. A NON STANDARD FAMILY OF PARTICLES

The particles that possibly emerge from the representation $G_{ns} \approx \Gamma_1 \oplus \Gamma_1 \oplus \Gamma_2$ ("*ns*" is for non-standard) have been ignored so far. We consider that $G_{ns}$ comes from a degeneracy breaking $\Gamma_2 \to \Gamma_1 \oplus \Gamma_1$ in $G_F \approx \Gamma_2 \oplus \Gamma_2$. Then a family associated with the representation $[\Gamma_1 \oplus \Gamma_1 \oplus \Gamma_2] \otimes [\Gamma_1 \oplus \Gamma_3]$ can be generated. This gives rise to a family comprised of six particles. We find four bosons on one hand

-$(\Gamma_1 \otimes \Gamma_1) \oplus (\Gamma_1 \otimes \Gamma_1)$: two scalar bosons

-$(\Gamma_1 \otimes \Gamma_3) \oplus (\Gamma_1 \otimes \Gamma_3)$: two vector bosons

and two fermions on the other hand

-$(\Gamma_2 \otimes \Gamma_1)$ : one lepton.

-and finally $(\Gamma_2 \otimes \Gamma_3)$ : one quark.

It is possible that one of these particles corresponds to the particle recently observed at the Large Hadron Collider (LHC). To understand which particle is really at stake we must, however, gain much more information regarding the properties of the found particle.

The two scalar bosons, if they do exist, behave as Higgs scalar particles and could be involved in the same sorts of reactions. If a Higgs boson is defined as a scalar boson the proposed particles are Higgs bosons. If a Higgs boson is defined as a particle that gives masses to otherwise zero mass particles the two particles are not Higgs bosons and we must find a mechanism that gives a left chirality to neutrinos. This problem is considered in section 8.

The lepton could give rise to a new type of neutrino. There are experimental evidences for the existence of a new type of neutrino, called a sterile neutrino whose (till controversial) existence has possibly been revealed in LSND [5] or Mini-Boone experiments [6].

Finally we can ignore the representation $\Gamma_1 \oplus \Gamma_1 \oplus \Gamma_1 \oplus \Gamma_1$ which seems not to have any physical consequences.

## 6. ANALYTICAL EXPRESSIONS OF PARTICLES MASSES

The definition of the bare mass $m_P^b$ implies that $\kappa_P^b > 0$. If $\kappa_P^b < 0$ one has a problem since the mass of the bare particle becomes imaginary. If one wants to keep the minimization principle, and to give nevertheless a physical meaning to negative eigenvalues, one is forced to interpret the negative eigenvalues as zero mass (or almost zero mass) particles. To make this point more precise we consider the set of states $\phi_\nu$ defined by $\Lambda_P(\phi_\nu) = 0$. The index "$\nu$" stands for neutrino as we shall see. This set forms a surface (a manifold) $\Sigma_\nu$ in the internal space of the world points that make a bare particle *P*. During the minimization process the state $\phi_P$ evolves so as to minimize $\Lambda_P(\phi_P)$ under the constraint $\phi_P^T \phi_P = 1$. The trajectory of $\phi_P$ possibly crosses $\Sigma_\nu$ at a point $\phi_P^\nu$. Let us then consider a state $\phi_P = \phi_P^\nu + d\phi_P$ . One has

$$\Lambda_P(\phi_P) \cong \Lambda_P(\phi_P^\nu) + \nabla\Lambda_P\big|_{\phi_P=\phi_P^\nu} d\phi_P = \nabla\Lambda_P\big|_{\phi_P=\phi_P^\nu} d\phi_P$$

$\phi_P$ has a physical meaning if $m^2 > 0$ that is if $\nabla\Lambda_P\big|_{\phi_P=\phi_P^\nu} d\phi_P > 0$. Crossing $\Sigma_\nu$, however, induces a sign change of $\nabla\Lambda_P\big|_{\phi_P=\phi_P^\nu} d\phi_P$. If $\phi_P$ has a physical meaning in one side of $\Sigma_\nu$ it loses this meaning on the other side. To a trajectory of state $\phi_P$, however, there corresponds a mirror trajectory of the anti-state $\overline{\phi_P}$. On this trajectory the sign of $\nabla\Lambda_P\big|_{\phi_P=\phi_P^\nu}$ is changed. Therefore while crossing $\Sigma_\nu$ a state $\phi_P$, attracted by $\Sigma_\nu$, keeps a physical meaning if it transforms into its anti-state $\overline{\phi_P}$ which is also attracted by $\Sigma_\nu$ . The trajectory of $\phi_P$ therefore stops on, or remains close to, $\Sigma_\nu$. Since $\Lambda_P(\phi_P^\nu) \cong 0$ the state $\phi_P = \phi_P^\nu$ corresponds to zero, or almost zero, mass particles.

The mass of a dressed particle is proportional to $\kappa_P$, which taking into account the effect of virtual bosons, is given by $\kappa_P^d = \kappa_P^b + \kappa_P^{vb}$ (*d* is for dressed, *b* is for bare and *vb* is for virtual bosons). If $\kappa_P^b < 0$ while $\kappa_P^d$ is positive, the mass $m_P^d$ of the dressed particle keeps a physical meaning which, in the framework of the present model, seems to be the case for the electron.

.

The Lagrangian operator of a particle, that is of a coupled pair of world points, one of bosonic type, the other of fermionic type, must keep the symmetry properties of the two components taken separately so as to preserve the organization of the particles of the Standard Model. This compels the Lagrangian to be a 16-dimensional operator given by

$$\Lambda_P = G_B \otimes 1^{(4)} + 1^{(4)} \otimes G_F + \zeta G_B \otimes G_F \qquad (8)$$

$I^{(4)}$ is the four dimensional unit matrix. The coupling parameter $\zeta$ is determined by the organization of interactions between the two world points that constitute a particle, more precisely by the relative numbers $n_+$ and $n_-$ of ferromagnetic and antiferromagnetic interactions that link the two points. For example $\zeta = (n_+ - n_-)/(n_+ + n_-)$.

The Lagrangian (8) has four eigenvalues. The first is associated with the down quark

$$\kappa_{down}^b = \kappa_{B2} + \kappa_{F2} + \zeta\kappa_{B2}\kappa_{F2} \qquad (9)$$

The second is associated with the up quark

$$\kappa_{up}^b = \kappa_{B2} + \kappa_{F1} + \zeta\kappa_{B2}\kappa_{F1} \qquad (10)$$

That is

$$\kappa_{down}^b = 1 + \left(\frac{1}{c} + \sqrt{3}\right) + \zeta\left(\frac{1}{c} + \sqrt{3}\right) \qquad (11)$$

and

$$\kappa_{up}^b = 1 + \left(\frac{1}{c} - \sqrt{3}\right) + \zeta\left(\frac{1}{c} - \sqrt{3}\right) \qquad (12)$$

According to eq. (6) with $K_s = K_Q$ the mass ratio between the two bare quarks is given by

$$m_{up}^b / m_{down}^b = \left(\kappa_{up}^b / \kappa_{down}^b\right)^{1/2} = \left(\frac{1 + \left(\frac{1}{c} - \sqrt{3}\right) + \zeta\left(\frac{1}{c} - \sqrt{3}\right)}{1 + \left(\frac{1}{c} + \sqrt{3}\right) + \zeta\left(\frac{1}{c} + \sqrt{3}\right)}\right)^{1/2} \qquad (13)$$

There is no possible calculation of bare lepton masses and one must take the influence of the polarization cloud of electroweak particles into account. The eigenvalues are then given by

$$\kappa_\nu^d = \kappa_{B1} + \kappa_{F1} + \zeta\kappa_{B1}\kappa_{F1} + \kappa_\nu^{vb} \qquad (14)$$

for the neutrino $\nu$ and

$$\kappa_{el}^d = \kappa_{B1} + \kappa_{F2} + \zeta\kappa_{B1}\kappa_{F2} + \kappa_{el}^{vb} \qquad (15)$$

for the electron. Explicitly one has

$$\kappa_{\nu}^{d} = -\frac{1}{c^2} + \left(\frac{1}{c} + \sqrt{3}\right) - \frac{1}{c^2}\zeta\left(\frac{1}{c} + \sqrt{3}\right) + \kappa_{\nu}^{vb} \tag{16}$$

and

$$\kappa_{el}^{d} = -\frac{1}{c^2} + \left(\frac{1}{c} - \sqrt{3}\right) - \frac{1}{c^2}\zeta\left(\frac{1}{c} - \sqrt{3}\right) + \kappa_{el}^{vb} \tag{17}$$

These formulae are now used to give some numerical results that are compared with experimental data.

## 7. NUMERICAL CALCULATIONS

There is no direct measurement of quark bare masses because it is not possible to observe separate quarks due to their confinement properties. The value of separate quark bare masses can only be obtained through various theoretical calculations that usually lead to different numerical values. The results for quarks bare masses are rather scattered. For example it has been proposed that $3\,\mathrm{MeV} < m_{up}^{b} < 8\,\mathrm{MeV}$ and $5\,\mathrm{MeV} < m_{down}^{b} < 10\,\mathrm{MeV}$ but calculations, carried out on lattices by C. Davies and al. [7] give values for the bare masses of light quarks that, the authors claim, are much more accurate than the previously published data. They find $m_{up}^{b} = 2.01 \pm 0.14\,\mathrm{MeV}$ and $m_{down}^{b} = 4.79 \pm 0.16\,\mathrm{MeV}$.

The ratio between bare quark masses is better determined than their absolute values and this is the only value that one can calculate in the present theory without appealing to free parameters. From the above data one has $m_{up}^{b} / m_{down}^{b} = 0.420 \pm 0.043$. It is worth noting that, contrary to quarks, the only known lepton masses are for dressed particles.

There are two parameters in eq.(13): c and $\zeta$. The constant c may be determined through considerations relating to electroweak interactions given in contribution [2]. Explicitily $c = \mathrm{tg}(\theta_W)$ where $\theta_W$ is the Weinberg angle. Then

$$1/c^2 = \left[1/\sin^2(\theta_W)\right] - 1$$

With the experimental value $\sin^2(\theta_W) = 0.231$ this yields $1/c = 1.824$ and the mass ratio eq. (13) becomes

$$m_{up}^{b} / m_{down}^{b} = \left( \frac{1 + \left(\frac{1}{c} - \sqrt{3}\right) + \zeta\left(\frac{1}{c} - \sqrt{3}\right)}{1 + \left(\frac{1}{c} + \sqrt{3}\right) + \zeta\left(\frac{1}{c} + \sqrt{3}\right)} \right)^{1/2} = \left( \frac{1.093 + 0.093\zeta}{4.557 + 3.557\zeta} \right)^{1/2} \tag{18}$$

We have no clue as regards an evaluation of the coupling parameter $\zeta$ except that the values $\zeta = 0$ and $\zeta = 1$ are forbidden, $\zeta = 0$ because the two world points that form the particle would not be coupled and $\zeta = 1$ because the coupling would be so strong that the two world points would be in the same state. A value $\zeta = 0.5$ could be a reasonable guess but we will see that the results are, in actual fact, weakly $\zeta$ dependent.

The dependence of the quark masses ratio upon the coupling parameter $\zeta$ (eq.18) is displayed in Fig.1. We observe that the results fit the Davies estimate, $0.38 < \zeta < 0.46$, on a wide range of $\zeta$, from $\zeta = 0.2$ to $\zeta = 1.0$. The average value $m_{up}^{b} / m_{down}^{b} = 0.420$ of Davies is obtained for $\zeta = 0.535$ close to the proposed value $\zeta = 0.5$. For $\zeta = 0.5$ (that is $n_{+} / n_{-} = 3$) one has $m_{up}^{b} / m_{down}^{b} = 0.424$. The mass of the bare up particle is given by $m_{up}^{b} = K_Q (\kappa_{up})^{1/2}$. With $m_{up}^{b} = 2.01\text{MeV}$ and $\kappa_{up} = 1.138$ the mass coefficient $K_Q$ is $K_Q = 1.884\text{MeV}$.

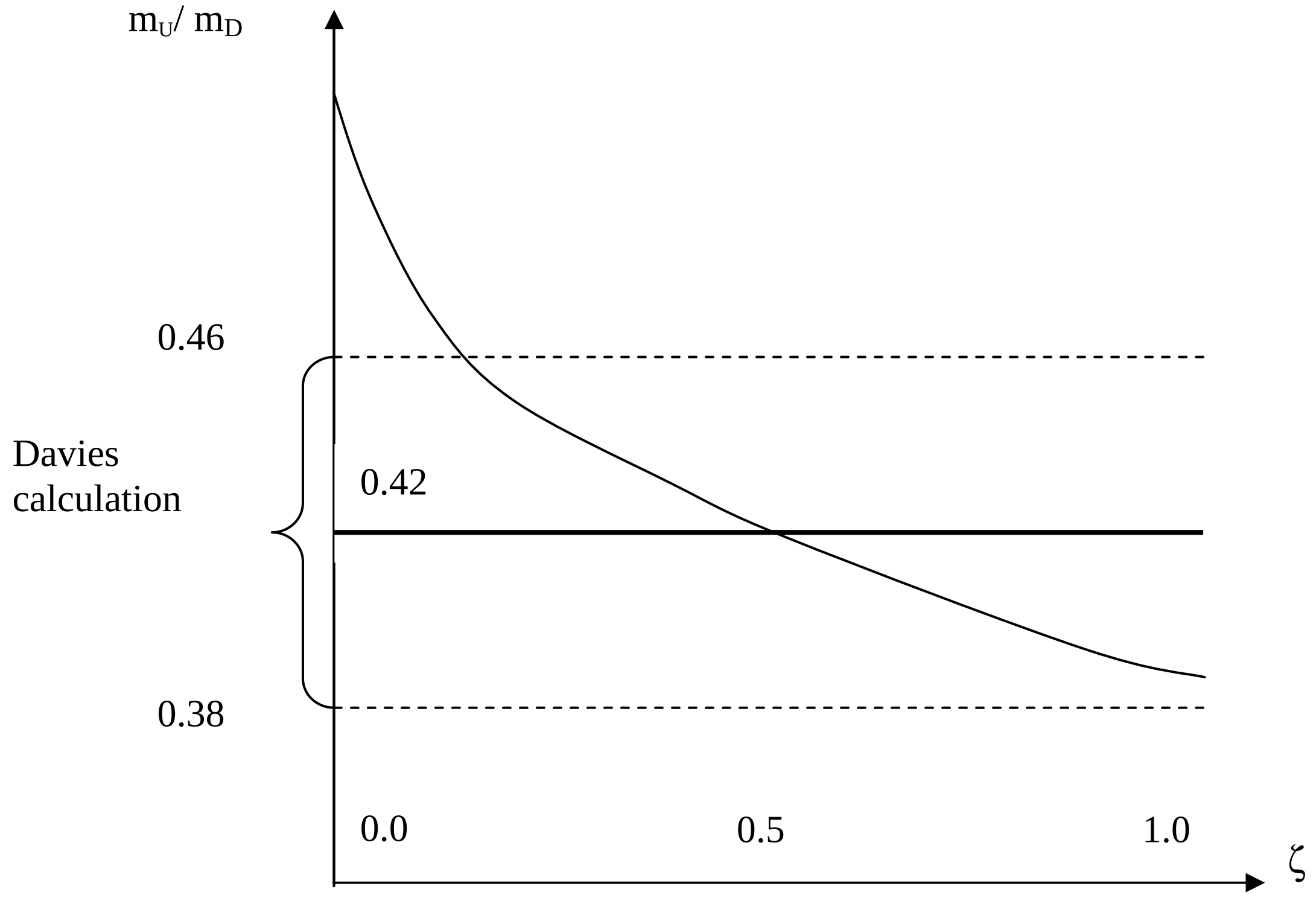

Fig-1: Evolution of the ratio of bare quark masses $m_{u}^{b} / m_{d}^{b}$ along the coupling parameter ζ

With the values adopted for c and $\zeta$ one has $\kappa_{\nu}^{d} = -5.690 + \kappa_{\nu}^{vb}$ and $\kappa_{el}^{d} = -3.442 + \kappa_{el}^{vb}$. $\kappa_{el}^{vb}$ must be positive and large enough for $\kappa_{el}^{d}$ to be positive. The calculation of the self energy $m_{el}^{vb}$ has not been carried out in the litterature. Assuming that $K_Q = K_L$, which certainly is a very crude approximation, the present analysis provides an order of magnitude of the expected result. From the electron mass $m_{el}^{d} = 0.51\text{MeV}$ one finds

$$m_{el}^{d} = 0.51 = K_Q \sqrt{\kappa_{el}^{vb} - 3.442}$$

That is $\kappa_{el}^{vb} = 3.71$. Then $m_{el}^{vb} = K_Q . \kappa_{el}^{vb} = 6.98\text{MeV}$

Since $\kappa_{el}^{vb} > \kappa_{\nu}^{vb}$ (the neutrino is neutral) one has $\kappa_{\nu}^{d} < 0$. Therefore the mass of the neutrino must vanish or, at most, remain very close to zero. In our interpretation indeed the state of a neutrino associated with negative eigenvalues oscillates from one side of the manifold $\Sigma_\nu$ to the other side and from the particle to its antiparticle. The mass of a neutrino is the mass it has in the positive side of $\Sigma_\nu$. Since the particle must never go far away from $\Sigma_\nu$ its mass is very small.

## 8. ON NEUTRINO'S LEFT HANDED CHIRALITY

The discovery (by Mrs Wu) that the parity symmetry is completely broken by weak interactions and that the electronic neutrino is left handed compelled the physicists to assume that the universe is, in actual fact, made of two different sorts of universes namely a left handed universe and a right handed universe. This hypothesis, in turn, leads to the conclusion that all fermions are zero mass particles. The well-known argument that supports the conclusion may be found in appendix B.

The eigenvalue associated with neutrinos are negative $\kappa_{\nu}^{d} < 0$ and the state $\phi_\nu$ of a neutrino must stay on or remain in the close vicinity of the manifold $\Sigma_\nu$, a surface on which the neutrino state can move freely until the Lagrangian $\Lambda(\phi_\nu) = \phi_\nu^T G_F \phi_\nu$ is minimized, that is until the neutrino state $\phi_\nu$ is parallel to a relevant eigenvector of $G_F$. $G_F$ can be written as

$$G_F = \frac{1}{c} I^{(4)} + \overline{G_F} = \frac{1}{c} I^{(4)} + \begin{pmatrix} 0 & \sigma_F \\ \sigma_F & 0 \end{pmatrix}$$

with

$$\sigma_F = \begin{pmatrix} 1 & 1+\mathrm{i} \\ 1-\mathrm{i} & -1 \end{pmatrix}$$

($\mathrm{i}^2 = -1$). The eigenstates of $G_F$ and those of $\overline{G_F}$ are identical. $\overline{G_F}$ has two eigenvalues. The eigenvalue $\sqrt{3}$ is associated with the neutrino and the eigenvalue $-\sqrt{3}$ is associated with the electron. Let us write the eigenstate of a neutrino as

$$\phi_\nu = \begin{pmatrix} \varphi_1 \\ \varphi_2 \\ \varphi_3 \\ \varphi_4 \end{pmatrix} = \begin{pmatrix} \phi_1 \\ \phi_2 \end{pmatrix}$$

For $\phi_\nu$ to be an eigenstate of $\overline{G_F}$ it is necessary that $\phi_1$ and $\phi_2$ are eigenstates of $\sigma_F$ with the same eigenvalue $\sqrt{3}$, corresponding to neutrinos, and $\phi_1$ and $\phi_2$ must be identical to one another within a phase factor $\exp(\mathrm{i}\eta)$ whence

$$\phi_\nu = \begin{pmatrix} \phi_1 \\ \exp(\mathrm{i}\eta)\phi_1 \end{pmatrix}$$

The Lagrangian of a neutrino writes

$$\Lambda(\phi_\nu)=\phi_\nu^T\overline{G_F}\phi_\nu=\left[\left(\phi_1^T,\exp(-\mathrm{i}\eta)\phi_1^T\right)\begin{pmatrix}0 & \mathrm{I}^{(2)}\sqrt{3}\\ \mathrm{I}^{(2)}\sqrt{3} & 0\end{pmatrix}\begin{pmatrix}\phi_1\\ \exp(\mathrm{i}\eta)\phi_1\end{pmatrix}\right]$$

that is

$$\Lambda(\phi_\nu)=\sqrt{3}(\exp(\mathrm{i}\eta)+\exp(-\mathrm{i}\eta))=2\sqrt{3}\cos(\eta)$$

$\Lambda(\phi_\nu)$ is minimum for $\eta=\pi$: $\Lambda(\phi_\nu)=-2\sqrt{3}$ and the eigenstate $\phi_\nu$ of a neutrino writes

$$\phi_\nu=\begin{pmatrix}\varphi_1\\ \varphi_2\\ -\varphi_1\\ -\varphi_2\end{pmatrix}$$

$\phi_\nu$ is an eigenvector of matrix $\gamma^5$ with eigenvalue -1 since

$$\gamma^5\phi_\nu=\begin{pmatrix} & & 1 & \\ & & & 1\\ 1 & & & \\ & 1 & & \end{pmatrix}\begin{pmatrix}\varphi_1\\ \varphi_2\\ -\varphi_1\\ -\varphi_2\end{pmatrix}=\begin{pmatrix}-\varphi_1\\ -\varphi_2\\ \varphi_1\\ \varphi_2\end{pmatrix}=-\begin{pmatrix}\varphi_1\\ \varphi_2\\ -\varphi_1\\ -\varphi_2\end{pmatrix}=-\phi_\nu$$

$\phi_\nu$ is therefore left-handed. In our approach the neutrino's left handed chirality is a consequence of the general properties of the universe which does not need to be shared into left handed and right handed universes. The same argument does not hold for the electronic state because $\kappa_e^d>0$ and the electron state trajectory does not cross the manifold $\Sigma_\nu$. Then the electron state is an eigenstate corresponding to the eigenvalue $\kappa_e^d$ which has no defined chirality.

The particles masses can be computed without appealing to a Higgs mechanism as we have seen. The Higgs field, identified with the polarization amplitude $\varphi_i$ ($\varphi_i=|\phi_i|$) of world point *i*, is still necessary not to give a mass to particles but simply to give them an existence. The Higgs field is also necessary to make sure that the photon mass is strictly zero as found by the GWS approach of electroweak interactions [2].

The model can however yield a value for the mass $m_H$ of Higgs particles if any. The polarization $\varphi$ minimizes the Landau expression [1]

$$F(\varphi)=\lambda\varphi^2+\mu\varphi^4$$

where

$$\lambda=\frac{n(1-bJ)}{2b}\ ;\ \mu=\frac{n}{12b}$$

*J* is the amplitude of binary interactions between the *n* cosmic bits of a world point and *b* is a parameter that materializes a cosmic disorder (somehow similar to a cosmic temperature though completely different in nature,). The minimum $\varphi_0$ of *F* is given by $\varphi_0^2=-\lambda/2\mu$. We expand *F* to second order around this minimum and find

$$F\cong-\lambda^2/4\mu-2\lambda(\varphi-\varphi_0)^2$$

which yields (see [1])

$$(m_H/\mathrm{c})^2 = -2\lambda = \frac{n(bJ-1)}{b}$$

In other respect one has $\hbar = l^*(n/2b)^{1/2}$ and therefore

$$m_H = \frac{\mathrm{c}\hbar}{l^*}(2(bJ-1))^{1/2}.$$

We now give to c its physical value. One has $l^*/\mathrm{c} = t^* = 1.6\times10^{-31}\mathrm{s}$ . With $bJ = 4.33$ and $\hbar = 6.6\times10^{-16}\mathrm{eV.s}$ one finds

$$m_H \cong 1.21\times10^4\mathrm{TeV/c^2}$$

a value that is far out of reach even for the most modern colliders.

## 9. REGGE TRAJECTORIES

The model presented here also provides a very simple explanation of Regge trajectories [8]. We have seen that the eigenvalue $\kappa_P$ of a particle, say that of a fermion, is related to the mass $m_P$ of the particle by the relation :

$$\kappa_P = (m_P^b)^2/K^2$$

Let us gather $n_f$ such fermions. The stability of the cluster is generally weak and the cluster, called a resonance (*R*), decays rapidly. The spins of the individual fermions adds up to a maximum spin $j = n_f\ \hbar/2$. Since the links between the individual fermions are weak their eigenvalues $\kappa_P$ approximately add up and $\kappa_R = (M_R)^2/K^2 = n_f\kappa_P = n_f\ (m_P)^2/K^2$ that is

$$(M_R)^2 \cong n_f(m_P)^2$$

Finally

$$j \cong (\hbar/2(m_P)^2)(M_R)^2$$

This linear dependence of the spin along the square of the mass $M_R$ is called a Regge trajectory.

## 10. THREE FAMILIES OF PARTICLES

As already argued the symmetry properties of bare particles are determined by the matrix $G_P$. The particles are created in high energy collisions experiments. In this type of experiments much energy is put down inside the core of a world point to such an amount that the value of cosmic temperature inside the core could be severely modified and, thereby, could modify its geometrical properties. We propose to look for the origin of families of fermions in these cosmic temperature perturbations.

We have seen, in [1], that the dimensionality d of space-time is given by $d = \mathrm{Int}(Jb)$ where Int is for integer part, *J* is the binary interaction between the cosmic bits and *b* is a measure of cosmic bits disorder that is of cosmic temperature [1]. The spatial dimension is $d_S = d-1$. A decrease of the cosmic temperature $b^{-1}$ results in an increase of the dimensionality *d*. When $b^{-1}$ decreases, the dimensionality of the core of the point of collision undergoes a series of changes.

(i) $1 > bJ > 0$ : $d = 0$ . The core of a world point is very "hot" and dimensionless

(ii) $2 > bJ > 1$ : $d = 1$ . The core has one time dimension and no space dimensions

(iii) $3 > bJ > 2$ : $d = 2$ . The core has one space dimension and one time dimension.

(iv) $4 > bJ > 3$ : $d = 3$ . The core has two space dimensions and one time dimension

(v) $5 > bJ > 4$ : $d = 4$ . This is the case for the ordinary space-time. The core has then three space dimensions and one time dimension.

The three domains (iii, iv and v) of cosmic temperatures would correspond to the three families of particles. The domain v ($d_S = 3$) would be the domain of the electron family, the domain iv ($d_S = 2$) that of the muon family and the domain iii ($d_S = 1$) that of the tau family. There is no room for any other family. There are three families of fermions because there are three space dimensions in our space-time.

## 12. CONCLUSION

The details of the organization of particles along the Standard model, their spins, charges and masses have been recovered in the framework of the discrete spaces model that we put forwardt (for electric charges see [2]). The neutrino chirality has been explained without appealing to a dichotomy between left and right universes. The introduction of this dichotomy is, in our opinion, too high a price to pay.

Our approach of physical phenomena implies a number of consequences:

i)-The concept of infinity does not belong to the realm of physics. Infinity would not be but a creation of mathematicians. We know that the observable universe is finite and therefore the infinitely large cannot be observed. We assume that, likewise, the infinitely small cannot be observed either.

ii)-The set of experimental results accumulated so far is enough to build a comprehensive theory of natural phenomena.

iii)-Many physicists believe that we cannot understand the laws of nature without appealing to complicated mathematics. Some physicists even believe that the necessary mathematics must be so complicated that they are outside the reach of a human brain. We do not share this prejudice. Notions of linear algebra and elements of group theory seem to be enough.

Finally let us list the main notions that have been introduced and that could be open to experimental observations.

a) First of all, we have introduced a metric limit $l^*$ (a cut-off) where both relativity and quantum mechanics lose their meanings. This solves the apparent incompatibility between these two theories. Exploring the metric limit implies energies of the order of $10^4$ TeV outside the reach of currently available colliders. The size $\rho l^*$ of the core of elementary bare particles, if large enough, could however correspond to sizes of the order of $10^2$ TeV that is of the order of $10^{-20}$ cm. Very large linear colliders, using the CLIC technology for example, could perhaps explore that range of energy.

b) We have defined a so called cosmic noise, a sort of temperature that describes the disorder of the most elementary physical systems (the cosmic bits) at a Planck scale. The cosmic noise, in our model, plays a central role in the organization and understanding of all natural phenomena. This noise could possibly be observed through phenomena relating to dark matter. In our model indeed the speed of light c is given by $c = 1/\sqrt{bJ-1}$. A variation of noise $b$ implies a modification of the metric tensor and, thus, phenomena that mimic the effects of dark matter. The modifications of the metric tensors, however, are different in the two interpretations. This point could perhaps be experimentally checked.

c) For quarks, our interpretation involves the representation $\Gamma_2 \otimes \Gamma_3$ implying gauge fields that must be invariant under the Lie group $SU(2) \times SU(3)$. Such a strong-weak interaction must modify the properties of strong forces, an effect that could perhaps be studied at the LHC. It is worth noting that no interaction is associated with the group $U(1) \times SU(2) \times SU(3)$ and therefore the Grand Unified Theory (GUT) is irrelevant in our approach.

d) The model has no need of a dichotomy of the universe between a right and a left universe. The left handedness character of a neutrino is a consequence of the present theory and no Higgs mechanism is necessary to give a mass to particles. The mass of the Higgs particle, if any, would be of the order of $10^4$ TeV.

e) The model opens the possibility to the existence of a new family of particles whose main properties are described in section V.

f) The physics of particles is deeply modified in strongly curved spaces as is the case in the vicinity of a black hole horizon (see [2]). In particular we have concluded that the strong interaction disappears in such an environment.

Acknowledgment: I would like to thank again Dr Ana Cabral for her very useful criticisms, suggestions and remarks.

## APPENDIX A: On the mass of bare particles

We consider a system of world points with $G_i = G_P$ and $G_{j \neq i} = G_0$. The world point $i$ is taken as the origin of coordinates and we require that its state $\phi_i$ is an eigenstates $\phi_S$ of $G_P$ $G_P \phi_S = \kappa_S \phi_S$. If the effects of the cloud of virtual particles are ignored altogether a time independent state $\psi_P^b$ of the system is the solution of the following Laplace equation:

$$\Delta \psi_P^b = \phi_s \delta(x) \qquad (A1)$$

"$s$" is for seed and "$b$" is for bare. The solution of (A1) is

$$\psi_P^b(r) = \phi_s (r / l^*)^{2-d_S} \qquad (A2)$$

where $r = |x|$ and $d_S$ is the spatial dimension of space-time. The perturbations caused by the seed on the core are copies of the seed polarization $\phi_s$ and therefore the symmetry of the bare particle is the same as the symmetry of the seed.

By using a first order perturbation approximation the perturbed metric matrices of the core may be written as perturbations of the vacuum metric tensor $G_j = G_0 + \eta_j G_P$, for $j \neq i$, with $|\eta_j| << 1$ and $\eta_j = 0$ for $|x_j - x_i| > \rho l*$.

Then the first order approximation of the eigenvalue, relative to vacuum, of the core of the bare particle *P* is given by:

$$\kappa_P^b = \kappa_s + \sum_{r(\neq 0)<\rho l*} \left(\phi_s^T \eta_S(r) G_P \phi_s\right)\left(\frac{r}{l*}\right)^{4-2d_S} = \kappa_s\left(1 + \sum_{r(\neq 0)<\rho l*} \eta_S(r)\left(\frac{r}{l*}\right)^{4-2d_S}\right) = K_s \kappa_s$$

$K_s = K_Q$ for quarks and $K_s = K_L$ for leptons.

## APPENDIX B: Diasappearance of mass terms in the Lagrangian of fermion fields

For completeness purposes let us briefly recall the (well known) arguments that lead to the conclusion that a dichotomy of the universe compels all particle masses to vanish. The mass term in the Lagrangian of a fermion field $\psi$ writes

$$m\psi^C\psi = m\psi^T\gamma^1\psi = m\left(\psi_R^C + \psi_L^C\right)\left(\psi_R + \psi_L\right)$$

where the Dirac matrix $\gamma^1$ has been defined in eq. (7).

We consider the right handed term $\psi_R^C\psi_R$. The projection operator $\left(1+\gamma^5\right)/2$ transforms a general state $\psi$ into a right handed state $\psi_R$

$$\psi_R = \frac{\left(1+\gamma^5\right)}{2}\psi$$

Since $\gamma^1\gamma^5 = -\gamma^5\gamma^1$ one also has

$$\psi_R^C = \psi^C\frac{\left(1-\gamma^5\right)}{2}$$

and thus

$$\psi_R^C\psi_R = \frac{1}{4}\psi^C\left(1-\gamma^5\right)\left(1+\gamma^5\right)\psi = \frac{1}{4}\psi^C\left(1-\left(\gamma^5\right)^2\right)\psi = 0$$

Likewise $\psi_L^C\psi_L = 0$. All other terms, such as $\psi_R^C\psi_L$, mix the chiralities and are therefore forbidden. Consequently there are no more mass terms left in the Lagrangian.
If there is no universe dichotomy, then this derivation no longer holds and the particles may have non zero masses.

References

[1] P. Peretto: Physics in discrete spaces (A): Space-time organization. ArXiv: 1012.5925v1

[2] P. Peretto: Physics in discrete spaces (B): Gauge and gravitation interactions. ArXiv: 1103.3815v1

[3] P.Peretto: Physics in discrete spaces (C): An interpretation of quantum states entanglement. ArXiv:1209.1853

[4] Y. Koide Phys. Rev. D **28**, 252 (1983)

[5] S. Dodelson, A. Melchiorri, A. Slosar (2006). Phys. Rev. Lett. **97**:04301. ArXiv 0511500.

[6] Miniboone,….: A combined oscillation analysis on the Mini-Boone excesses. ArXiv1207-4809

[7] C. Davies and al. : *Science Now* (2 April 2010)
Q. Mason, H.D. Trottier, R. Horgan, C.T.H. Davies, G. P. Lepage: ArXiv: hep-ph/0511.160v1 (2005)

[8] R. Penrose, G.A. Sparling, S.T. Tsou: *Journ. Phys. A Mat. Gen.* 11 231 (1978)
P. D. B. Collins: *An Introduction to Regge Physics and High Energy Physics* Cambridge University Press (1977)